\documentclass[referee]{raa}            

\usepackage{graphicx,times}             
\usepackage[noend]{algpseudocode}
\usepackage{algorithmicx,algorithm}
\usepackage{amsmath}	
\usepackage{amssymb}	

\usepackage{natbib}
\bibpunct{(}{)}{;}{a}{}{,}
\usepackage[pagebackref=true]{hyperref}
\usepackage{array}
\usepackage[a4paper, left=2.5cm, right=2.5cm, top=2.5cm, bottom=2.5cm]{geometry}

\begin{document}

  \title{Dealing with the data imbalance problem on pulsar candidates sifting based on feature selection
}

   \volnopage{Vol.0 (20xx) No.0, 000--000}      
   \setcounter{page}{1}          

   \author{Haitao Lin  
      \inst{1}
   \and Xiangru Li
      \inst{2}
   }

   \institute{School of Mathematics and Statistics,
Hanshan Normal University,
Chaozhou 521000, China\\
        \and
             School of Computer Science, South China Normal University, Guangzhou 510631, China; {\it xiangru.li@gmail.com}\\
\vs\no
   {\small Received 20xx month day; accepted 20xx month day}}

\abstract{ Pulsar detection has become an active research topic in radio astronomy recently. One of the essential procedures for pulsar detection is pulsar candidate sifting (PCS), a procedure of finding out the potential pulsar signals in a survey. However, pulsar candidates are always class-imbalanced, as most candidates are non-pulsars such as RFI and only a tiny part of them are from real pulsars. Class imbalance has greatly damaged the performance of machine learning (ML) models, resulting in a heavy cost as some real pulsars are misjudged.
To deal with the problem, techniques of choosing relevant features to discriminate pulsars from non-pulsars are focused on, which is known as {\itshape feature selection}. Feature selection is a process of selecting a subset of the most relevant features from a feature pool. The distinguishing features between pulsars and non-pulsars can significantly improve the performance of the classifier even if the data are highly imbalanced.
In this work, an algorithm of feature selection called {\itshape K-fold Relief-Greedy} algorithm (KFRG) is designed. KFRG is a two-stage algorithm. In the first stage, it filters out some irrelevant features according to their K-fold Relief scores, while in the second stage, it removes the redundant features and selects the most relevant features by a forward greedy search strategy. Experiments on the dataset of the High Time Resolution Universe survey verified that ML models based on KFRG are capable for PCS, correctly separating pulsars from non-pulsars even if the candidates are highly class-imbalanced.
}

   \authorrunning{H.-T. Lin \& X. -R. Li }            
   \titlerunning{Pulsar candidates sifting based on feature selection}  

   \maketitle

%
%
\section{Introduction}           
\label{sect:intro}
Pulsars are highly magnetized, rotating, compact stars that emit beams of electromagnetic radiation out of their magnetic poles. They are observed as signals with short and regular rotation periods when their beams are received by the earth.
The study of pulsars is of great significance to promote the development of astronomy, astrophysics, general relativity, and other fields.
As a remarkable laboratory, it can be used for the research of the detection of gravitational wave \citep{taylor1994binary}, the observation of the interstellar medium \citep[ISM;][]{han2004spatial}, the conjecture of dark matter \citep{baghram2011prospects} and other research fields. Therefore, many pulsar surveys (projects) have been carried on or ongoing to search for more new pulsars. These pulsar surveys have produced massive observation data in the form of pulsar candidates. For example, the number of pulsar candidates from the Parkes Multibeam Pulsar Survey \citep[PMPS;][]{manchester2001parkes} is about 8 million; the High Time Resolution {Universe} Pulsar Survey
\citep[HTRU;][]{keith2010high,levin2013high} has
returned 4.3 million candidates \citep{morello2014spinn}; the Low-Frequency Tied-Array All-Sky survey \citep[LOTAAS;][]{van2013lofar} has accumulated 3 million candidates, etc.
With the development of modern radio telescopes, such as the Five-hundred-meter Aperture Spherical radio Telescope \citep[FAST; ][]{nan2006five,nan2011five,nan2016fast} and Square Kilometre Array \citep[SKA; ][]{smits2009pulsar}, the amount of pulsar candidates increases exponentially. However,
of this vast amount of candidates, only a small part of these candidates are from real pulsars, while others are radio frequency interferences (RFI) or other kinds of noises
\citep{keith2010high}. Thus, one essential process of pulsar search is to separate the real pulsar signals from non-pulsar ones, which is known as pulsar candidate sifting (PCS).


Recently, quite a few machine learning (ML) methods have been applied to PCS.
They are mainly divided into two types according to their inputs----models based on artificial features and models based on image-driven approaches.

Artificial features are designed in accordance with the different nature between pulsars and non-pulsars.
 These features were extracted by their physical background (we called empirical features) or statistical characteristics (statistic features) and can be clearly explained. Typically,
\cite{eatough2010selection} first extracted 12 empirical features from candidates as inputs of an Artificial Neural Network \citep[ANN;][]{haykin1994neural, hastie2005elements} model for PCS. The ANN model was experimented in PMPS survey \citep{manchester2001parkes} and achieved a recall rate of 93\% (recall rate is a performance measure defined as a ratio between the number of the successfully predicted pulsars and the total number of the real pulsars). And Bates et~al. \cite{bates2012high} constructed another ANN with 22 features as inputs in HTRU-Medlat \citep{keith2010high} and achieved a recall of 85\%. To improve the performance of PCS, Morello et~al. \cite{morello2014spinn} empirically designed 6 empirical features to build a model called Straightforward Pulsar Identification using Neural Networks (SPINN), which even achieved both a high recall of 100\% and a low false positive rate.
Then, a purpose-built tree-based model called Gaussian Hellinger Very Fast Decision Tree (GHVFDT) \citep{lyon2014hellinger} was applied to PCS with 8 newly designed features. These features are statistics computed
from both the folded profile and the dispersion measure (DM) searching curve (defined in Section \ref{sec:dataset}). They are evaluated, using the joint mutual information criterion to helps identify relevant features.
 Later,
 Tan et~al. \cite{tan2017ensemble} pointed out that the GHVFDT based on these 8 features is insensitive to pulsars with wide integrated profiles. They therefore proposed eight other new features and built an ensemble classifier with five different decision trees to improve the performance of the detection.

 As for image-driven PCS models, they are based on deep learning networks, where deep features were extracted from their diagnostic plots (Fig. \ref{pulsarfig}). \cite{zhu2014searching} first proposed the pulsar imaged-based classification system (PICS), whose inputs are four mainly diagnostic plots, i.e., sub-integration, sub-band, the folded signal and the DM-curve ( their definitions can be referred to Section \ref{sec:dataset}). \cite{wang2019pulsar} then improved the PICS and designed a PICS-ResNet model which is composed of two Residual Neural Networks (ResNets), two Support Vector Machines (SVMs), and one Logistic Regression (LR). \cite{guo2019pulsar} used a combination of a deep convolutional generative adversarial network (DCGAN) and a support vector machine (SVM) to apply to the HTRU Medlat and PMPS surveys. Then they raised a model by combining the DCGAN and MLP neural networks trained with pseudo inverse learning auto encoder (PILAE) algorithm, achieving excellent results on class-imbalanced data sets (\cite{mahmoud2021learning}).
Recently, \cite{XIAOFEI2021364} designed a 14-layer deep residual network for PCS, using an over-sampling technique to adjust the imbalance ratio of the training data. The experiments on HTRU achieved both a high precision and a 100\% recall.
\cite{XIAOFEI2021364}
 As far as intelligent identification is concerned, deep learning methods have shown great significant application in the PCS. More related works can be referred to \cite{zhang2019pulsar, guo2019pulsar, zeng2020concat, xiao2020pulsar, lin2020pulsar}. Although these models showed an advantage in performance, they failed to quantify significantly difference between pulsars and non-pulsars since the deep features extracted from them were inexplicable or incomprehensible.

One of the greatest challenges for in ML is the class imbalance problem \cite{japkowicz2000learning}, where the distribution of instances with labels is skewed. In the case of a binary classification problem, class imbalance implies that the number of one class is far less than the number of the other class in a data set. We refer to these two categories as the majority and the minority, respectively. The ratio between the total number of the majority and that of the minority is called imbalance ratio (IR). A machine classifier with high IR tends to judge an unknown item as a major class, resulting in low recall rates.
For instance, HTRU dataset is a highly class-imbalanced set, as the number of non-pulsar signals is close to 90,000 while the number of pulsar signals is only 1196.
To address the class imbalance problem, oversampling methods were implemented before training. For instance, Morello et al. \cite{morello2014spinn} balanced their training set by randomly oversampling to 4:1 ratio of non-pulsars to pulsars. Bethapudi et al. \citep{bethapudi2018separation} and Devine et al. \cite{devine2016detection} adopted the Synthetic Minority Over-sampling Technique \citep[SMOTE;][]{chawla2002smote} to produce more positive samples and raise the recall of the models. SMOTE is one of the most commonly used oversampling methods to handle the imbalanced data distribution problem. It generates virtual instances by linear interpolation for the minority class. These instances are generated by randomly choosing one or more of the k-nearest neighbors for each example in the minority class. After the oversampling process, the data are reconstructed to be class-balanced. However, this is not reflective of the real problem faced and is an artefact of data processing since these generative instances are virtual, random and not from the real world.

 In our work, instead of raising the performance of PCS models by balancing the training data of candidates, we improve the pulsar accuracy of the models in perspective of the features representation, which is called \textbf{feature selection} or \textbf{variable selection} \citep{tang2014feature} in ML terminology. Exactly, feature selection is the process of selecting a subset of relevant features from the feature candidate pool. Feature selection is necessary in the data preprocessing stage, as some of the features may be redundant or irrelevant. These redundant or irrelevant features will decrease the performance of the sifting model. A well-designed feature selection algorithm will significantly improve the predictive ability of the ML sifting model. For the above considerations, an algorithm of feature selection called {\itshape K-fold Relief-Greedy (KFRG)}
is proposed in this work. KFRG is a purpose-built two-stage algorithm: the first stage is to filter out some irrelevant features from the candidate features by Relief score, while the second stage is to select the most relevant features in a greedy way.
To verify the effectiveness of KFRG for PCS, several typical ML classifiers are evaluated, including C4.5 \citep{quinlan2014c4},  Adaboost \citep{freund1997decision}, Gradient Boosting Classification \citep[GBC;][]{moller2016photometric},  XGboost \citep{chen2016xgboost} etc. Our experiments were performed on the public data of HTRU {\citep {keith2010high}}.

The article is arranged as follows. Section 2 gave the description of the HTRU dateset as well as some related works. In Section 3, as many as 22 artificial features were introduced, including 6 empirical features from \cite{morello2014spinn}, 8 statistical features designed by \cite{lyon2016fifty} and 8 additional statistical features proposed by \cite{tan2017ensemble}. These features were collected to be selected in the next progress. In Section 4, KFRG as an algorithm of feature selection for PCS was proposed. Experiments based on KFRG were carried on HTRU survey data in Section 5, and the discussion and conclusion was made in Section 6.

\section{Data and Preliminary Works}

\subsection{Pulsar candidates and the HTRU dataset}
\label{sec:dataset}

\begin{figure}
\centering
\includegraphics[width=12 cm]{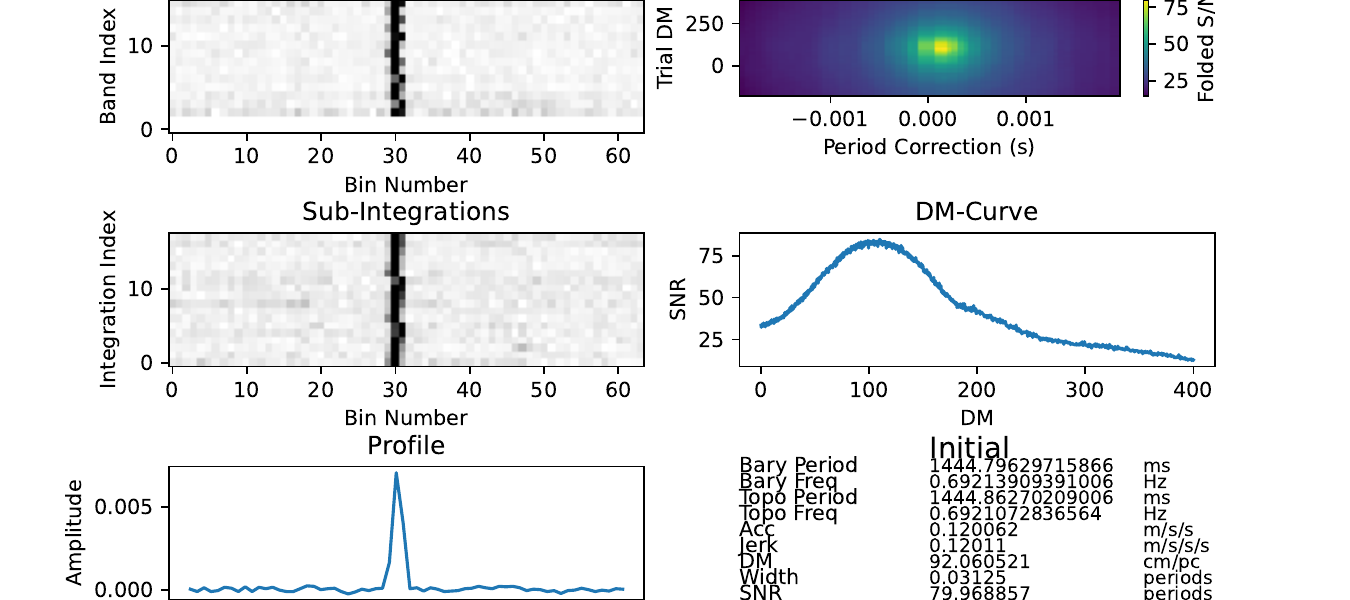}
\caption{Diagnostic plots and summary statistical characters of a pulsar candidate. }\label{pulsarfig}
\end{figure}

 A pulsar candidate is originally a piece of signal from the receiver of a radio telescope during the observation time.
 Most commonly, it is processed by PulsaR Exploration and Search TOolkit \citep[PRESTO;][]{ransom2001new}, a typical software for pulsar search and analysis. Then the candidate is represented using a series of physical values and a series of diagnostic plots as Fig. \ref{pulsarfig} shows. On the left, the plots from top to bottom are :
 a sub-band plot, a sub-integration plot and a folded profile of the signal.
 A sub-band plot displays the pulse in different bands of observed frequencies; a sub-integration plot shows the pulse in the time domain; a folded profile is the folded signal of its sub-bands by frequency or the folded signal of its sub-integrations by period. On the right are two plots. One is a grid searching plot for dispersion measure (DM) and period. The other is a DM-searching curve.
 It is known that DM measures the number of electrons which the pulsar’s signal travel through from the source to the Earth. However, real DM is unknown and should be obtained by trials. Then a grid searching plot on the right top of Fig. \ref{pulsarfig} records the change of SNR as the trial DMs and the trial periods vary. A DM searching curve on the right middle describes the relationship between trial DM and its corresponding SNR, and the peak of the curve implies the most likely value of DM.

Our work is conducted on HTRU survey.
The HTRU data set is observed with two Observatories, the Parkes radio telescope in Australia and the Effelsberg 7-beam system in Germany.
HTRU is an ambitious (6000 hours) project to search for pulsars and fast transients in the entire sky, which is split into three areas:
Low-latitude (covers $\pm$ 3.5$^\circ$), Mid-latitude (Medlat; covers $\pm$ 15$^\circ$) and High-latitude (covers the remaining sky $<$ 10$^\circ$).
The pipeline searched for pulsar signals with DMs from 0 to 400 $pc\cdot cm^-3$ (DM is often quoted in the units of $pc\cdot cm^-3$, which makes it easy to estimate the distance between a given pulsar and the earth), and also performed an acceleration search between  -50 to +50 $m\cdot s^{-2}$.

HTRU data set was publicly released by \cite{morello2014spinn}) and available online
\footnote{{http://astronomy.swin.edu.au/$\sim$vmorello/}} available.
It consists of 1196 real pulsar candidates and 89,995 non-pulsars, which are highly class-imbalanced, as only a tiny fraction of the candidates are from real pulsar signals (Table \ref{sample}).

\begin{table}
        \begin{center}
        \caption { The description of our experimental data. About 70\% of the total candidates are selected as training sets, others are test samples. The imbalance ratios of HTRU is as high as 75:1.}\label{sample}
        \begin{tabular}{llrrrrr}
        \hline \noalign{\smallskip}
         Data  &Part&Amount & Pulsar & Non-pulsar & IR   \\
        \hline
        \noalign{\smallskip}
        HTRU&Total(100\%)& 91,192 & 1196& 89,996 &75:1\\
        &Training(70\%)& 63,834 & 837& 62,997 &75:1\\
          &Test(30\%)& 27,357 &359&26,999  &75:1\\
          \noalign{\smallskip}
        \hline
        \end{tabular}
        \end{center}
\end{table}

\subsection{Machine Learning Classifiers}
PCS can be described as a binary supervised classification issue in ML. Supervised learning \citep{mitchell1997machine} is an ML task of learning a function that maps instances to their labels. Particularly,
 a classifier of PCS aims to learn a function mapping features of the pulsar candidates to their categories----pulsar or non-pulsar. To evaluate the effectiveness of selected features, the performance of classifiers should be estimated.

In our work, feature selection algorithms are evaluated by 7 classifiers. Among them, Decision Tree \citep[DT;][]{quinlan2014c4}, Logistic Regression \citep[LR;][]{hosmer2013applied} and Support Vector Machine \citep[SVM;][]{suykens1999least} are normal classifiers, while Adaptive boosting \citep[Adaboost;][]{freund1997decision}, Gradient Boosting Classification
\citep[GBDT;][]{moller2016photometric}, eXtreme Gradient boosting \citep[XGboost;][]{chen2016xgboost} and Random Forest \citep[RF;][]{liaw2002classification} are ensemble learning classifiers.
The principles of these classifiers are different and representative. For example, typical DT classier is based on information gain ratio while SVM tries to find the best hyperplane which represents the largest separation between the two classes. Ensemble methods \citep{dietterich2002ensemble} use multiple weak classifiers such as DT to obtain a strong classifier.
Interested readers can refer to their references for details \citep{mitchell1997machine, mohri2018foundations}.

\subsection{Performance metric}
\label{Sec:PerformanceMetric}
To evaluate the performance of a classifier on class-imbalanced data, typically on pulsar candidates data, four most relevant metrics are given. There are the \textit{Recall} rate, the \textit{Precision} rate, the \textit{$F_1$ score} and the \textit{False Positive Rate} ($FPR$). They can be expressed by \textit{True Positives} (TP), \textit{True Negatives} (TN), \textit{False Positives} (FP) and \textit{False Negatives} (FN).

In binary classification, $recall$, defined by ${TP}/{(TP+FN)}$, where ${TP}$ denotes the amount of true pulsars predicted and ${TP+FN}$ the total amount of true pulsars. It measures how many pulsars could be correctly predicted pulsars from all the real pulsars. $Precision$, defined by ${TP}/{(TP+FP)}$, measures how many true pulsars would be predicted correctly out of the candidates predicted tasks. However, $recall$ and $precision$ compose the relationship that opposite to each other, as it is probable to increase one at the cost of reducing the other. Therefore, $F_1$ score, defined as the harmonic mean of $recall$ and $precision$, i.e., ${(2\cdot Precision \cdot Recall)}/{(Precision +Recall)}$, is a trade-off between them.
As for $FPR$, it measures the ratio of mislabelled non-pulsars out of all the non-pulsar candidates by ${FP}/{(TN+FP)}$. It can be inferred by $recall$ and $precision$. Therefore, we just need to focus on the $recall$, the $precision$ and the $F_1$ score of a classifier.

\section{Feature pool}
Before feature selection, a set of candidate features should be collected to be further selected, which is called a feature pool.
Feature selection algorithm will be implemented on this pool to output a feature subset of better representation.
Considering that some of the candidate features are trivial for PCS, several guidelines for a feature pool were discussed in this section.
\subsection{Guidelines of candidate features}\label{guidelines}
To extract some robust and useful features, guidelines of feature design were proposed by PCS researchers.
 \cite{morello2014spinn} gave several suggestions, such as "ensuring complete robustness to noisy data" in order to "exploit properly in the low-SNR regime", "limiting the number of features" to deny "the curse of dimensions".  \cite{lyon2016fifty} suggested that features should be designed to maximize the separation between positive and negative candidates, reducing the impact of class imbalance.

Based on suggestions from \cite{morello2014spinn} and \cite{lyon2016fifty}, guidelines for candidate features in a feature pool were summarized as follows. Candidate features:
\begin{itemize}
      \item[i.] should be distinguishable enough between pulsars and non-pulsars. A distinguishable feature will greatly improve the performance of the classifier.
      \item[ii.] should be diversified. Considering the diversity of the feature source, both empirical features and statistical features should be included in the feature pool.
      \item [iii.] should be full-covered.
      Features should be extracted from the mainly diagnostic images, especially the sub-integration plot, the sub-band plot, the folded profile and the DM searching curve.
      \item [iv.] can be easily extracted and calculated.
      \item [v.] should be controlled in a moderate total number. If the total number is small, some relevant features may be missed; if it is large, it will enlarge the computing cost of the feature selection algorithms.
\end{itemize}
\subsection{Candidate features in our work}
Following the guidelines in Section \ref{guidelines},
twenty-two features have been collected (Table \ref{fea1to22}), which are candidate features from \cite{morello2014spinn, lyon2016fifty} and \cite{tan2017ensemble}.
Among them, six features were defined by Morello et al.  \citep{morello2014spinn} to build the \textit{SPINN} model, eight statistical features were proposed by Lyon et al. \cite{lyon2016fifty} as inputs to \textit{GHVFDT}, and eight additional features were introduced by Tan et al.\cite{tan2017ensemble} to developed an ensemble classifier comprised of five different decision trees. Details of these features were described in Table \ref{fea1to22}.


\begin{table}
\begin{center}
        \caption {Notations and definitions of 22 candidate features in our work. Features with ID M1-M6 were defined by Morello et al.(2014); Features with ID L1-L8 were created by Lyon et al.(2016); Features with ID T1-T8 were were defined by Tan et al.(2017). }\label{fea1to22}
    \begin{tabular}{llllllllll}
        \hline\noalign{\smallskip}
         ID &Feature &Description \\
        \noalign{\smallskip}
        \hline\noalign{\smallskip}
        M1&  $log(SNR)$          & Log of the signal-to-noise of the pulse profile                       \\
        \noalign{\smallskip}
        M2&$D_{eq}$& Intrinsic equivalent duty cycle of the pulse profile  \\
        \noalign{\smallskip}
        M3&$Log(P/DM)$&      Log of the ratio between period and $\texttt{DM}$        \\
        \noalign{\smallskip}
        M4&$V_{DM}$&       Validity of optimized $\texttt{DM}$                        \\
        \noalign{\smallskip}
        M5&$\chi_{(SNR)}$  &   Persistence of the signal in the time domain\\
        M6&$D_{RMS}$& RMS between folded profile and sub-integration       \\
        \noalign{\smallskip}
        L1&  $Pf_{\mu}$ & Mean of the folded profile                  \\
        L2&$Pf_{\sigma}$   & Standard deviation  of the folded profile                   \\
        L3&$Pf_{k}$ & Kurtosis of the folded profile                  \\
        L4&$Pf_{s}$ & Skewness of the folded profile                   \\
        L5&  $DM_{\mu}$ & Mean of the $\texttt{DM}$ curve                 \\
        L6&$DM_{\sigma}$   & Standard deviation of $\texttt{DM}$ curve                  \\
        L7&$DM_{k}$ & Kurtosis of $\texttt{DM}$ curve                  \\
        L8&$DM_{s}$ & Skewness of  $\texttt{DM}$ curve                  \\
        T1&$SubbandCorr_{\mu}$ & Mean of the $SubbandCorr ^1$                \\
        T2&$SubbandCorr_{\sigma}$   & Standard deviation of $SubbandCorr$                  \\
        T3&$SubbandCorr_{k}$ & Kurtosis of $SubbandCorr$                 \\
        T4&$SubbandCorr_{s}$ & Skewness of $SubbandCorr$                  \\
        T5&$SubintCorr_{\mu}$ & Mean of the $SubintCorr^2$                 \\
        T6&$SubintCorr_{\sigma}$   & Standard deviation of $SubintCorr$                  \\
        T7&$SubintCorr_{k}$ & Kurtosis of $SubintCorr$                 \\
        T8&$SubintCorr_{s}$ & Skewness of $SubintCorr$                  \\

        \noalign{\smallskip}\hline

    \end{tabular}

    \begin{tabular}{lll}
    Note:&\\
             1.$SubbandCorr$: & A vector of correlation coefficient between\\
             &  sub-band and the folded profile. \\
        2.$SubintCorr$: & A vector of correlation coefficient between \\
        & sub-integration and the folded profile. \\
    \end{tabular}
\end{center}
\end{table}

\begin{figure}
  \centering
  \includegraphics[width=14 cm]{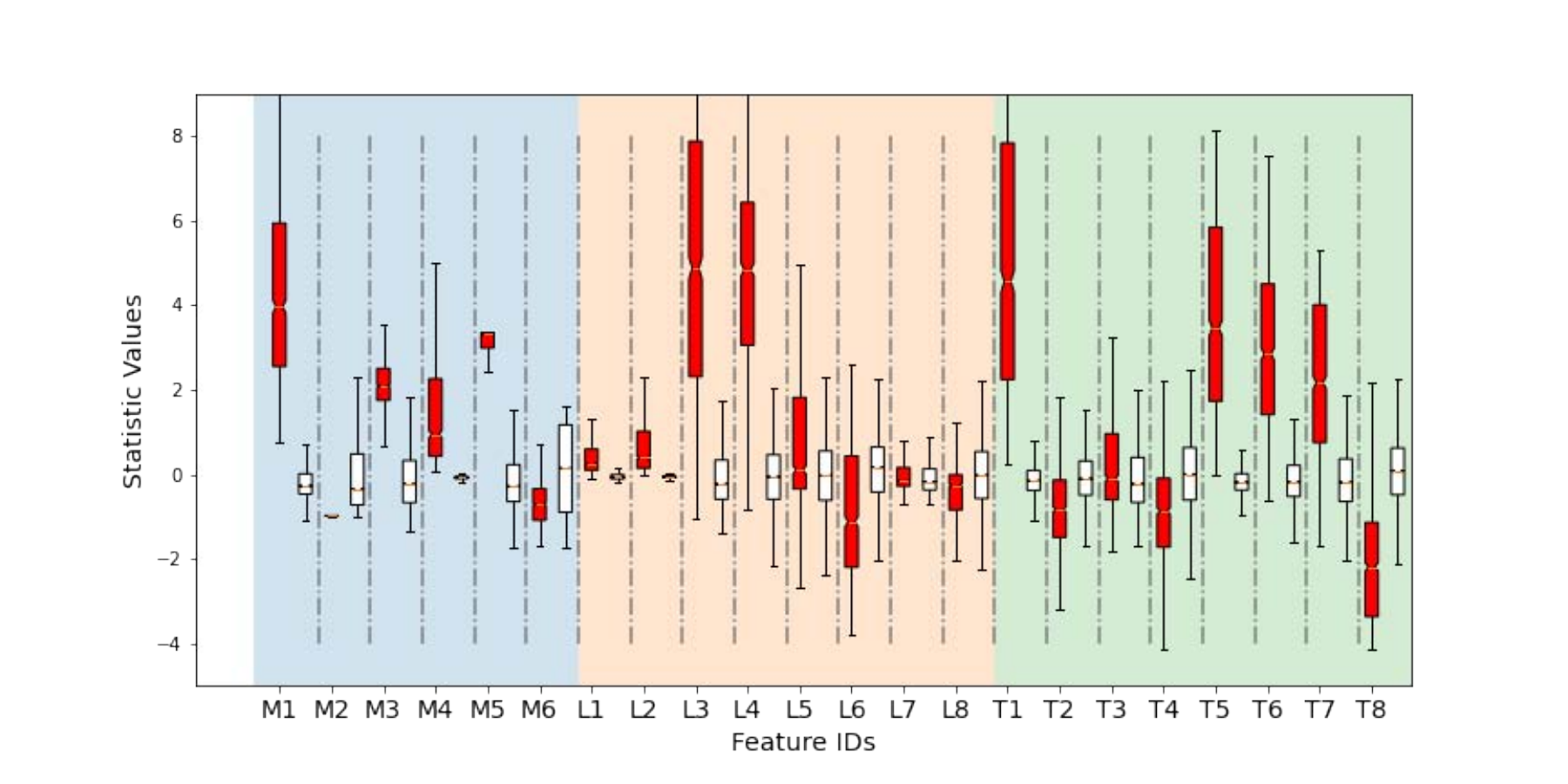}\\
  \caption{Box plots of features on HTRU.  For each feature there are two distinct box plots, of which the red box describes the distribution for known pulsars in the feature, while the white one is for non-pulsars.}\label{boxplot_HTRU}
\end{figure}

To demonstrate the discriminating capabilities of these features, one statistic approach is to show the distributions of pulsars and non-pulsars from each feature by box plots, which can graphically demonstrate the locality, spread, and skewness groups of the features.
Figure \ref{boxplot_HTRU} gives box plots of our candidate features on HTRU. There are two box plots per feature. The red boxes describe the feature distribution for known pulsars, while the white ones are for non-pulsars mainly consisting of RFI.
Note that the data of each feature was scaled by z-score, with the mean zero and the standard deviation one. The resulting z-score measures the number of standard deviations that a given data point is from the mean. Generally, the less the overlap of the red box and white box in a feature, the better separability of the feature. However, the usefulness of features according to their box plots are only on a visual level. Measurable investigation of these features will be given in the next section.

\section{Feature Selection Algorithms}

\subsection{Motivation}
Feature selection algorithm is a search technique for a feature subset from a feature pool. Irrelevant or redundant features not only increase the calculation efficiency of ML models but also damaged their performances. Better selected features
can be helpful when facing data-imbalanced problem \cite{2004SPECIAL} and some practical algorithms have been proposed for two-class imbalanced data problem \citep{2013Feature,2014Feature}.

There are mainly three categories of feature selection algorithms: filters, wrappers, and embedded methods.
Filter methods use a proxy measure to score a feature subset. Commonly, they include the mutual information \citep{shannon1948mathematical}, the point-biserial correlation coefficient \citep{gupta1960point}, and Relief \citep{urbanowicz2018relief} score.
Wrapper methods score feature subsets using a predictive model.
Common methods include grid search, greedy \citep{black2012greedy} and recursive feature elimination \citep{kira1992practical}. An embedded feature selection method is a machine learning algorithm that returns a model using a limited number of features.

Our proposed algorithm of feature selection will combine the idea of a filter called Relief with a wrapper method named greedy.
On one hand, PCS as a mission of binary supervised classification, is highly dependant on the constructive features between pulsars and non-pulsars. Thus, methods of filter are preferably considered as they measure the relation between features and labels. In fact, measures by filters are able to capture the usefulness of the feature subset only based on the data, which are independent of any classifier. Relief algorithm is one of the best measures of filters when compared with other filter methods. It weights features and avoids the problem of high computation cost in combinatorial search. Thus, our proposed approach of feature selection for PCS is a Relief-based algorithm.
On the other hand, the selected features according to their Relief scores may be redundant. It involves that more than two features with high Relief scores but they are strongly correlated, since one relevant feature may be redundant in the presence of another relevant feature \citep{guyon2003introduction}. To remove these redundant features, it follows the evolution of Relief scores to implement a greedy technique.

However, Greedy or Relief for feature selection has some shortages.
Although Relief score is able to filter out some irrelevant features, it could not detect the redundant ones, which implies that if two features share the same information in terms of correlation measure, both of them are very likely judged as relevant or irrelevant.
As for Greedy, it can be utilized to reduce the number of features.
Its computational cost increases in a quadratic way as the number of features increases, which makes the computer unaffordable.

Based on the considerations above, we combine Relief with the Greedy algorithm to
propose an KFRG algorithm of feature selection for PCS.
In the first stage, it filters out some irrelevant features according to their Relief scores, while in the second stage, it removes the redundant features and selects the most relevant features by a forward greedy search strategy.
Experimental investigations on HTRU showed that it improved the performance of most classifiers and achieved high both recall rate and precision (Section \ref{Sec:experiments}).

\subsection{Relief Algorithm}\label{Sec:reliefs}
Relief (\cite{kira1992practical}) is a filter algorithm of feature selection which is notably sensitive to feature interactions.
It calculates a feature score for each feature which can then be applied to rank and select top scoring features for feature selection. Alternatively, these scores may be applied as feature weights to guide downstream modeling. Relief is able to detect conditional dependencies between features and their labels (pulsars and non-pulsars) and provide a unified view on the feature estimation in regression and classification. It is described as Formula \eqref{Eq:relief}. The greater the Relief score of a feature, then the more distinguishable the feature is, and corresponding features are more likely to be selected.

Let $D=\{(X_i,y_i) | X_i=(x_i^1,x_i^2,\cdots,x_i^d), i\in I\}$ denote a dataset, where $y_i$ is the label of $X_i$ and $x_i^j$ the $j$th component of $X_i$.  Denote $x_{i,nh}^j$ the $j$th feature of its nearest instance whose label is the same as that of $X_i$ (a 'nearest hit'), while $x_{i,nm}^j$  the $j$th feature of its nearest instance whose label is  different from that of $X_i$ (a 'nearest miss'). Then Relief score of the $j$th feature (denoted as $\delta  ^j $) is defined as
\begin{equation}\label{Eq:relief}
\delta  ^j =\frac{1}{|I|}\sum \limits_{i\in I}(-diff(x_i^j,x_{i,nh}^j)^2+diff(x_i^j,x_{i,nm}^j)^2),
\end{equation}
where $diff$ represents the difference of two components.
$
diff(x_a^j,x_{b}^j)=
\begin{cases}
0, & y_a=y_b\\
1, & y_a\neq y_b
\end{cases}
$ if the $j$th feature is discrete, while
$
diff(x_a^j,x_{b}^j)=|x_a^j-x_{b}^j|
$ if the $j$th feature is continuous.


\begin{algorithm}
\caption{Relief Algorithm}\label{algorithm1}
\hspace*{0.02in}
{\bf Input:}\\
$D$;  \qquad \quad      \#Training dataset\\
$S$;    \ \qquad \quad  \#Features pool \\
$K$;   \qquad \quad  \#A preset threshold\\
\hspace*{0.02in} {\bf Output:}\\
$\Delta$.        \qquad \quad  Set of selected features\\
----------------------------------------------------------------------------
\begin{algorithmic}[1]
\State  $\Delta=[\ ]$ \qquad \# Initialization of the chosen features
\For{$j = 1: |S|$}
\State Compute $\delta  ^j$ by Equation \eqref{Eq:relief}
\If{$\delta  ^j > \delta_0$}
\State $\Delta= \Delta \cup \{j\}$
\EndIf
\EndFor
\State \Return $\Delta$
\end{algorithmic}
\end{algorithm}

\subsection{Greedy Algorithm}\label{Sec:greedy}

A greedy algorithm is an algorithm that follows the problem-solving heuristic of making the locally optimal choice at each stage \cite{black2012greedy}. Greedy strategy does not usually produce an optimal solution. Nonetheless, a greedy heuristic method may yield locally optimal solution that approximates the global optimal solution in a reasonable amount of time.

In our algorithm, the objective function (target) is to maximize the $F_1$ score of a classifier,
 and thus our greedy is designed to choose the best feature step by step from the rest of the feature pool. Here, ``the best feature candidate" is the feature which contributes best to increase of the $F_1$ score of the classifier. Accordingly, the stopping criterion of our greedy in the iteration procedures is that the $F_1$ score does not increase anymore, or the maximum number of the selected features is more than a preset threshold $Maxlen$.

 {$Maxlen$ is a hyperparameter to control the maximum number of the selected features. On one hand, $Maxlen$ should be large enough to enable as many as possible candidate features. A small $Maxlen$ may miss some relative features and result in low performance.
 On the other hand, as $Maxlen$ increases, so dose the computational complexity} \citep{goldreich2008computational}.
 {Computational complexity is an important part of an algorithm design, as it gives useful information about the amount of resources required to run it. In fact, the computational complexity of Greedy can be expressed as $O(Maxlen^2)\times O(\mathfrak{L})$ according to Algorithm \ref{algorithm2}, where the notation $O(Maxlen^2)$ means the run time or space requirements grow as the square of $Maxlen$ grows, while $O(\mathfrak{L})$ represents the computational complexity of $\mathfrak{L}$ which relies on both the choice of a classifier and the size of the input (feature). Thus, a large $Maxlen$ implies a large computing cost.
 To evaluate a fit $Maxlen$, experiments with $Maxlen$ ranging from 2 to 10 were carried out. Experimental investigation shows that as $Maxlen$ increases, the average performance metrics improve rapidly at first, while they keeps in a similar level when $Maxlen$ is more than 8. The average recall and precision keep around 97.2\% and 98.0\% for $Maxlen$ larger than 8 as Table \ref{Tab:comparing} shows. Considering both the computational complexity and the performance, $Maxlen$ is set to be 8.
}

Following the description above, our \textbf{Greedy Algorithm} of feature selection can be described as {Algorithm} \ref{algorithm2}.
\begin{algorithm}[htbp]
\caption{Greedy Algorithm}\label{algorithm2}
\hspace*{0.02in}
{\bf Input:}\\
$D$;  \qquad \qquad      \# Training dataset\\
$S$;    \ \qquad \qquad   \# Features pool \\
$\mathfrak{L}$; \ \qquad \qquad \# Classifier\\
$Maxlen$;   \ \quad  \# Maximum number of selected features\\
\hspace*{0.02in} {\bf Output:}\\
$S_{ch}$.        \ \ \qquad \quad  \# Set of selected features\\
----------------------------------------------------------------------------
\begin{algorithmic}[1]
\State  $S_{ch}=[\ ]$;        \qquad \# Initialization of the chosen features
\State $S_{un}=S$;  \qquad \# Initialization of the unchosen features
\State $F_1=0$;       \qquad\   \# Initialization value of $F_1$ score
\While {$|S_{ch}|<MaxLen$}
    \For {$j \in S_{un}$}
    \State $\mathfrak{L}_{j} = \mathfrak{L}(S_{ch} \cup\{j\})$ \quad {\small{\# add $j$ to a classifier}}
    \EndFor
    \State $j^*=\arg\max\limits_{j\in S_{un} } {F_1(\mathfrak{L}_{j})}$
       \quad  {\small{\# choose the best feature}}
    \State $F_1^* = F_1(\mathfrak{L}(S_{ch} \cup\{j^*\})$

\If{$F_1^*>F_1$}
\State $F_1=F_1^*$
        \State $S_{un}=S_{un} \setminus \{j^*\}$
        \State $S_{ch}=S_{ch}\cup \{j^*\}$ \quad {\small{\# add $j$ to a classifier}}
\Else
        \State break
\EndIf
\EndWhile
\State \Return $S_{ch}$
\end{algorithmic}
\end{algorithm}

\subsection{KFRG Algorithm}\label{Sec:RGA}
Relief is easy to operate, but it is not very satisfied in some class-imbalanced scenarios, since it may underestimate those features of high discriminative ability in minority, and ignores the sparse distributional property of minority class samples \citep{Yuanyu2019A}. That is, a feature with high Relief score pays more attention to the non-pulsars which are the majority class and hard to identify the pulsars which are the minority class and thus more promising pulsars will miss.

To overcome these flaws, the K-fold Relief algorithm (KFR) was designed.
The key improvement of KFR is to balance the data by recycling the minority and sampling from the majority. Firstly, split the training data into minority samples and majority ones. Then, produce K disjoint subsets from the majority samples randomly, and merge each subset with the minority into K new data sets, each of which is relatively balanced with the same minority samples. Here, K is a preset integer which is normally the ratio of the majority classes to minority classes in training data set. Finally, calculate the mean of Relief scores of each set. KFR is able to promote the importance of the minority classes for the estimation of relevant features.

Combining KFR with Greedy algorithm, we get KFRG. KFRG is a two-stage algorithm: the first stage aims to remove some irrelevant features from the candidate features according to their Relief score, while the second stage is designed to select the most relevant features in a greedy way. It can be described in Algorithm \ref{algorithm3}.

\begin{algorithm} [htbp]
\caption{KFRG Algorithm}\label{algorithm3}
\hspace*{0.02in}
{\bf Input:}\\
$D$;  \qquad \quad      \#Training data\\
$S$;    \ \qquad \qquad    \#Feature pool \\
$\mathfrak{L}$; \ \qquad \qquad \#A classifier\\
$Maxlen$;   \ \quad  \#Maximum number of selected features\\
\hspace*{0.02in} {\bf Output:}\\
$S^*$.        \ \ \qquad \quad  \#Set of selected features\\
----------------------------------------------------------------------------
\begin{algorithmic}[1]
\State Split the training data $D$ into minority class $Mi$ (pulsars) and majority class $Ma$ (non-pulsars);
\State Divide $Ma$ into K disjoint subsets, $Ma=\cup^{K}_{k=1}(Ma)_{k}$;
    \For {$k=1:K$}
    \State Obtain the Relief scores of $S$ on the data set $(Ma)_{k} \cup Mi$ by Algorithm \ref{algorithm1} (Relief).
    \EndFor
\State Calculate the mean of Relief scores of $S$ to choose the a feature subset $\Delta$.
\State Implement Algorithm \ref{algorithm2} (Greedy) on $\Delta$ by setting the maximum number of selected features as $Maxlen$.
\State \Return $S^*$
\end{algorithmic}
\end{algorithm}

\section{Experiments and Analysis}\label{Sec:experiments}

{ In this section, experiments based on KFRG were implemented on HTRU.
Firstly, selected features by KFR and KFRG were calculated, respectively.
Then, to demonstrate the improvement of KFRG, ablation study was given to see the contribution of the component to the KFRG. Afterwards, comparative experiments with different feature groups were carried out to verify the effectiveness of KFRG. Finally, comparative experiments between our proposed KFRG and oversampling approachs were given, and their advantages and disadvantages of each were discussed.}

{ \subsection{Results of KFG and KFRG}}

\begin{figure}[htbp]
\centering
\includegraphics[width=14cm]{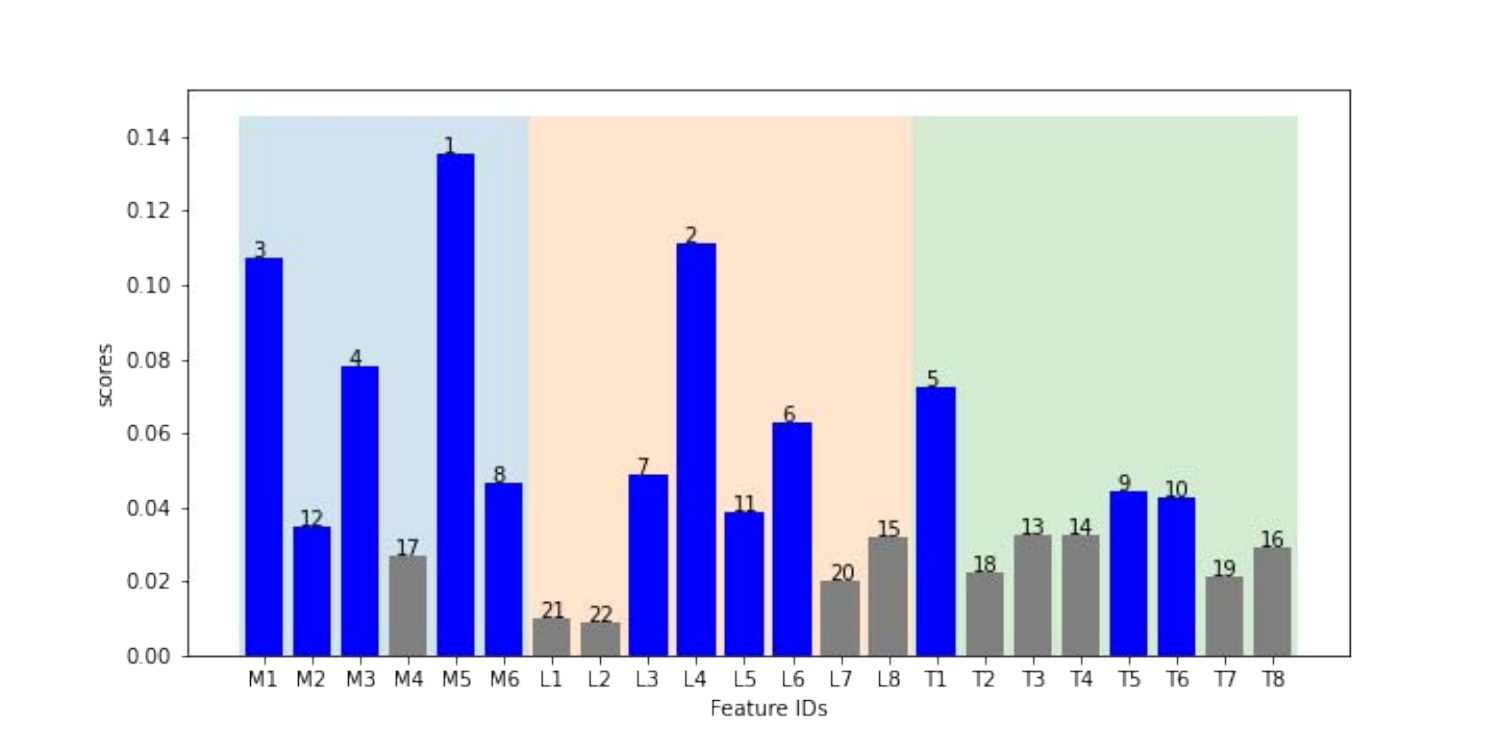} 
\caption{A bar graph of K-fold Relief scores of 22 features and their ranks on HTRU.  A higher Relief score implies a more important feature. With a preset ratio, all features are divided into relevant ones (blue bars) and others (gray bars).}\label{score}
\end{figure}

{\subsubsection{Selected features based on KFR scores}}
\label{Sec:ReliefAnalyze}
The KFR algorithm is the first stage of Algorithm~\ref{algorithm3}, which outputs the mean of Relief scores of the K-fold training sets. The Relief scores standing for the weights of features were calculated and shown by a bar graph in Fig. \ref{score} for HTRU, where both the scores and their ranks were given. To select the more relevant features, a preset threshold will keep the features with higher scores. Here, a ratio of 0.618 is preset to ensure that the number of selected features is more than half of the total number of the candidate features. That is, about 12 features out of 22 are considered to be much relevant ones (blue bars) and 10 others (gray bars) are less relevant.

{\subsubsection{Selected features based on KFRG}}
\label{Sec:RG}

Based on KFRG (Algorithm \ref{algorithm3}), the selected features as well as their size were obtained and the results were given in Table \ref{Tab:greedy}.

\begin{table}
\begin{center}
\caption{Selected features based on the KFRG algorithm.}\label{Tab:greedy}
\setlength{\tabcolsep}{1.5mm}{
\begin{tabular}{lllllllll}
\hline
   Classifier & Selected Feature &Size  \\
\hline
{DT}		       &	[ M5, L4, T5]    &3             \\
LR & [ M2, M5, L4, L5, L6, T1, T5, T6 ] &8     \\
SVM	    	&   [ M2, M3, M5, L4, T5 ]  &5       \\
Adaboost      &[ M2, M3, M5, L4, L3 ]   &5         \\
GBDT	   	&[ M3, M5, L3, L4, T1 ]     &5       \\
XGBoost	   	&[ M5, M6, L4, T1, T5 ]     &5       \\
RF	        &[ M2, M3, M5, L4, T1 ]     &5          \\
\hline
\end{tabular}
}
\end{center}
\end{table}
\begin{itemize}
  \item[\textbullet] The dimension of features is greatly reduced. Most of the features in the feature pool were removed after KFRG algorithm. The dimension of the selected features is cut down from 22 to less than 8. Some classifiers even need only 3 features to build their models.
  \item[\textbullet] The selected features and their sizes vary with the classifiers. For example, only three features M5, L4, T5 were finally chosen for DT while five features M2, M3, M5, L4, T1 were left for RF.
  \item[\textbullet] Features M5 and L4 are frequently used in all of the classifiers. It is shown that features M5 from \cite{morello2014spinn} and L4 from \cite{lyon2016fifty} are frequently selected. Further discussion is given in Section \ref{Sec:discussion}.

\end{itemize}

{\subsection{Ablation study of KFRG}}

{KFRG feature selection algorithm is an improvement of the Relief algorithm by combining K-folded Relief (KFR) with a greedy algorithm. To demonstrate the effectiveness of KFRG, an ablation study of stack mode is given. Step by step, the performance metrics of all the classifiers were calculated from Relief to KFR, and finally, to KFRG.}

\begin{table}[htbp]
\begin{center}
\caption{ Ablation study of KFRG. Performance metrics of several classifiers based on Relief, K-folded Relief scores and our proposed KFRG algorithm were given, including their recall (Rec), prcision (Pre), $F_1$ scores and false positive rate ($FPR$).
}\label{Tab:ablation}
    \begin{tabular}{lllllllllllllllllll}
    \hline
      {Algorithm}  & \multicolumn{4}{c}{{Relief}} && \multicolumn{4}{c}{{KFR}}&& \multicolumn{4}{c}{{KFRG}}\\
      \cline{2-5}
      \cline{7-10}
      \cline{12-15}
      & Rec &Pre&$F_1$& {$FPR$}&& {Rec} &{Pre}&{$F_1$} &{$FPR$}&& {Rec} &{Pre}&{$F_1$}&{$FPR$}\\
      \hline

{DT}      &{\bf{97.3}}&{94.7}&{96.0}&{0.05}&&{96.7}&{97.3}&{97.0}&{0.05}&&{
96.4}&{\bf{98.0}}&{\bf{97.2}}&{\bf{0.03}}\\
{LR}      &{86.3}&{93.0}&{89.7}&{0.08}&&{93.0}&{\bf{96.7}}&{94.7}&{\bf{0.04}}&&{
\bf{96.0}}&{95.8}&{\bf{95.9}}&{0.06}\\
{SVM}     &{90.0}&{\bf{99.3}}&{94.3}&{\bf{0.02}}&&{92.3}&{98.7}&{95.3}&{\bf{0.02}}&&{
\bf{96.4}}&{98.0}&{\bf{97.2}}&{0.03}\\
{Adaboost}&{98.3}&{96.7}&{97.3}&{\bf{0.03}}&&{98.0}&{\bf{97.7}}&{97.7}&{\bf{0.03}}&&{
\bf{98.3}} &{97.5}&{\bf{97.9}}&{\bf{0.03}}\\
{GBDT}    &{97.0}&{99.0}&{98.3}&{\bf{0.02}}&&{97.7}&{\bf{99.0}}&{98.3}&{\bf{0.02}}&&{
\bf{97.8}}&{98.9}&{\bf{98.3}}&{\bf{0.02}}\\
{XGBoost} &{98.0}&{98.3}&{98.1}&{0.03}&&{\bf{98.3}}&{98.7}&{\bf{98.3}}&{\bf{0.02}}&&{
97.8}&{\bf{98.9}}&{\bf{98.3}}&{\bf{0.02}}\\
{RF}      &{97.3}&{97.3}&{97.3}&{0.04}&&{96.7}&{\bf{99.7}}&{\bf{98.0}}&{\bf{0.02}}&&{
\bf{97.5}}&{98.6}&{\bf{98.0}}&{\bf{0.02}}\\
\hline
{Mean} &{94.8}&{96.9}&{95.8}&{0.039}&&{96.1}&{\bf{98.2}}&{97.0}&{\bf{0.03}}&&{
 \textbf{97.2}}&{98.0}&{\textbf{97.5}}&{\textbf{0.03}}\\
\hline
    \end{tabular}\\[2pt]
  \end{center}
\end{table}

\begin{figure}[htbp]
  \centering
  \includegraphics[width=14 cm]{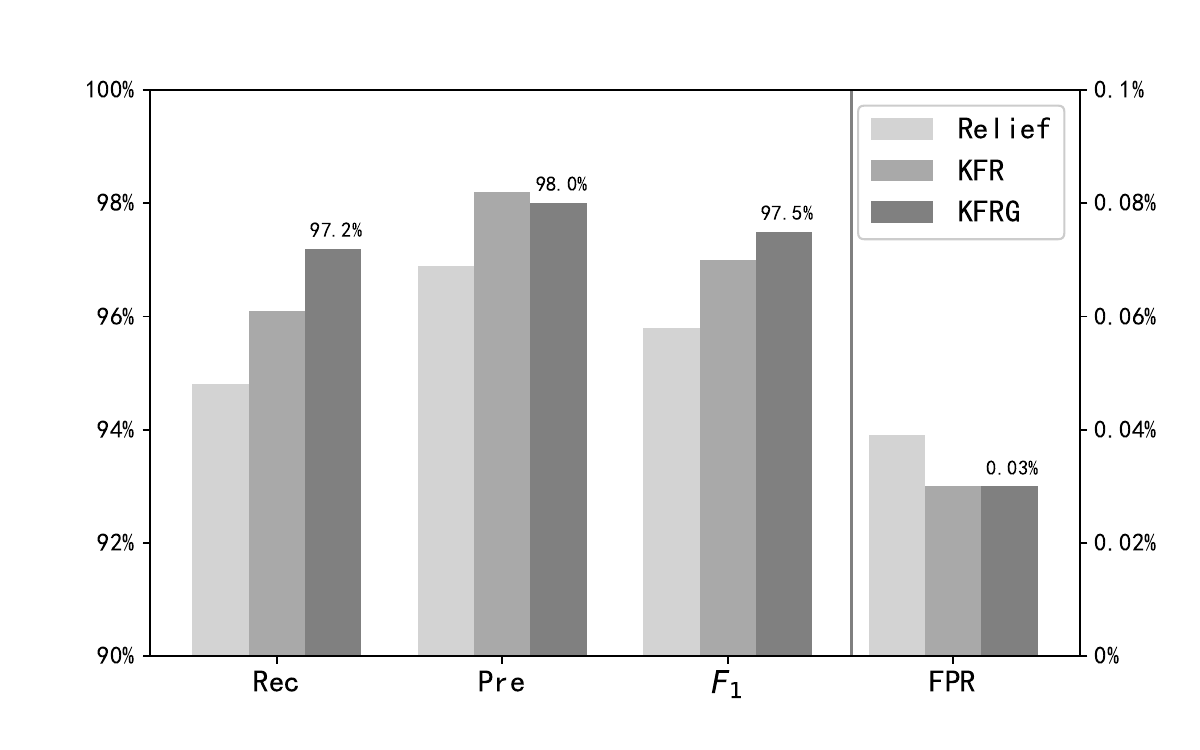}\\
  \caption{The average performance metrics of Relief, KFR and KFRG.}\label{Ablation_plot}
\end{figure}

{Table \ref{Tab:ablation} gives the numerical calculation of recall, precision, $F_1$ score and false positive rate of each classifiers with three different feature select algorithms----Relief, KFR and KFRG, while
Figure \ref{Ablation_plot} plots their averaged performance metrics.
It shows that KFR performs better than Relief, and KFRG performs best of all.
For one thing, KFR has better recall rates, better precision and lower FPRs for all classifiers than the original Relief techniques. For example, the recall raises from 94.8\% to 96.1\%. These improvements come from the step of $K$-fold Relief operation, as KFR is designed for the imbalance problem.
For another, KFRG keeps a precision rate as high as KFR, and raises the recall rate by 0.9\%. Further, KFRG achieves a best $F_1$ score of 97.5\%. These improvements come from the step of Greedy as it aims to maximize the $F_1$ score by removing the redundant features.
}


{\subsection{Comparison of performance with different features}}

To verify the effectiveness of KFRG, comparative experiments with different feature groups were carried out, where three subsets of the feature pool were considered as the inputs of the classifiers, including features from \cite{morello2014spinn}, features from \cite{lyon2016fifty}, and features from KFRG. The performance metrics of recall (Rec), precision (Pre), $F_1$ score and false positive rate (FPR) were given in Table \ref{Tab:comparing}.

\begin{table}
  \centering
  \begin{center}
  \caption{Performance metrics based on different features, including M1-M6 from Morello et al., L1-L8 from Lyon et al. and the selected features by our proposed KFRG algorithm (Table \ref{Tab:greedy}), where Rec, Pre, $F_1$ and $FPR$ stand for performance metrics of recall, precision, $F_1$ score and false positive rate.}
  \label{Tab:comparing}
  \begin{minipage}[t]{1.2\textwidth}
    \begin{tabular}{lllllllllllllllllll}
    \hline
      {Features}  & \multicolumn{4}{c}{M1-M6} && \multicolumn{4}{c}{L1-L8}&& \multicolumn{4}{c}{Selected features by KFRG}\\
      \cline{2-5}
      \cline{7-10}
      \cline{12-15}
      & Rec &Pre& $F_1$ & $FPR$&& Rec &Pre&$F_1$&$FPR$&& Rec &Pre&$F_1$&$FPR$\\
      \hline
{DT}      & 89.7 & 95.4 & 92.5 & 0.06 &&
 84.1 & 93.7 & 88.6 & 0.08 &&
\bf{96.4} & \bf{98.0} & \bf{97.2} & \bf{0.03} \\
LR      & 85.0 & 90.1 & 87.6 & 0.12 && 82.1 & 90.5 & 86.1 & 0.12 &&
\bf{96.0} & \bf{95.8} & \bf{95.9} & \bf{0.06}\\
SVM     & 92.7 & 93.6 & 93.1 & 0.08 && 75.4 & 96.6 & 84.7 & 0.04 &&
\bf{96.4} & \bf{98.0} & \bf{97.2} & \bf{0.03}\\
Adaboost& 86.0 & 94.9 & 90.2 & 0.06 && 84.4 & 91.7 & 87.9 & 0.10 &&
\bf{98.3} & \bf{97.5} & \bf{97.9} & \bf{0.03}\\
GBDT    & 92.4 & 93.9 & 93.1 & 0.08 && 89.0 & 94.7 & 91.8 & 0.07 &&
\bf{97.8} & \bf{98.9} & \bf{98.3} & \bf{0.02}\\
XGBoost & 89.4 & 94.7 & 92.0 & 0.07 && 85.0 & 96.2 & 90.3 & 0.04 &&
\bf{97.8} & \bf{98.9} & \bf{98.3} & \bf{0.02}\\
RF      & 88.4 & 96.0 & 92.0 & 0.05 && 81.4 & 95.3 & 87.8 & 0.05 &&
\bf{97.5} & \bf{98.6} & \bf{98.0} & \bf{0.02}\\
\hline
Mean & {89.1} & {94.1} & {91.5} &{0.074} &&{83.1} &{94.1} &{88.2} &{0.071} &&
 {\textbf{97.2}} &{\textbf{98.0}} &{\textbf{97.5}} &{\textbf{0.03}}\\

\hline
    \end{tabular}\\[2pt]
  \end{minipage}
  \end{center}
\end{table}

The experimental results in Table \ref{Tab:comparing} are summarized as follows.
\begin{itemize}

  \item[\textbullet] Both the recall and the precision have significantly improved.\\ The average recall of the classifiers is $97.2\%$, and most of the classifiers achieve recall rates ranging from 96\% to 99\% after KFRG algorithm, which implies that most of the real pulsar signals were well detected after feature selection. For instance, the recall rate and the precision rate of DT are respectively 89.7\% and 95.4\%  with features M1-M6, while they are up to 96.4\% and 98\% based on KFRG. Also, the average precision is as large as $98.0\%$ based on KFRG, which has increased by an average of 3.9\% compared with M1-M6 and L1-L8.

  \item[\textbullet] The $FPR$ of the classifiers is reduced to 0.05\% in average.\\ Most of the classifiers achieve a $FPR$ less than $0.05\%$. A low $FPR$ of a classifier implies that the selected features are very sensitive in excluding non-pulsars.
  \item[\textbullet]  F1 scores based on selected features have increased. \\ The best $F_1$ is 98.3\% in our experiment in the following cases: using five selected features [M3, M5, L3, L4, T1], GBDT achieved a recall of 97.8\% and a precision of 98.9\% ; Using another five selected features [M5, M6, L4, T1, T5], XGBoost also achieved a high $F_1$ score of 98.3\%, which is as good as the GBDT classifier. A better $F_1$ score implies that both $recall$ and $precision$ increase since $F_1$ score is the harmonic mean of $recall$ and $precision$. In other words, more potential pulsar signals are correctly recognized and fewer non-pulsar signals are misjudged in these cases.
\end{itemize}

{\subsection{Comparison of performance with different data-balancing techniques}}

As feature selection of KFRG alleviates the imbalance problem on ML, the performances based on KFGR were compared with some other widely-used data-balancing techniques of over sampling, including randomly oversampling, SMOTE \citep{chawla2002smote}, Borderline SMOTE \citep{2005Borderline} and ADASYN \citep{he2008adasyn}.
Furthermore, we even implement KFRG-SMOTE method which is a combination of the proposed KFRG and SMOTE technique.
We evaluated the metrics of each algorithm on the different classifiers, and then took the mean of the performance of each classifier on each evaluation statistic to compare between feature selection metrics and oversampling metrics (Table \ref{Tab:comparison_oversample}).

\begin{table}[htbp]
\begin{center}
\caption{{Performance metrics with different data-balancing techniques, including randomly oversampling, SMOTE, Borderline SMOTE, ADASYN, our proposed KFRG algorithm, and KFRG-SMOTE which is a combination of the proposed KFRG with SMOTE. Rec, Pre, $F_1$ and $FPR$ stand for performance metrics of recall, precision, $F_1$ score and false positive rate.}}\label{Tab:comparison_oversample}
\setlength{\tabcolsep}{1.5mm}{
\begin{tabular}{lllllllll}
\hline
   {Method} & {Rec} &{Pre}& {$F_1$} & {$FPR$} \\
\hline
	{Randomly oversampling}   &{92.2}& {90.8}& {91.5}& {0.11}\\
	{SMOTE}                   &{97.3}& {88.3}& {92.6}& {0.15}\\
	{Borderline SMOTE}        &{97.0}& {89.0}& {92.8}& {0.14}\\
	{ADASYN}                  &{97.2}& {91.8}& {94.4}& {0.10}\\
	{KFRG}                    &{97.2}  & {\bf{98.0}} &{\bf{97.5}} &{\bf{0.03}}   \\
	{KFRG-SMOTE }             &{\bf{97.8}}& {97.3}& {\bf{97.5}}& {0.04}\\
\hline
\end{tabular}
}
\end{center}
\end{table}

{It shows that KFRG offers better performance than oversampling techniques such as SMOTE, Borderline SMOTE and ADASYN, and randomly oversampling performs worst among them. KFRG achieve a similar recall rate but an improved precision and a lower FPR, which imply that the classifiers on KFRG is very strict in the criteria for classifying the candidates as pulsars and only a few of non-pulsars were misjudged. Moveover, the performance of KFRG-SMOTE is as good as KFGR, which share a high $F_1$ scores of 97.5\% and whose FPRs both range at a low level between 0.03\% to 0.04\%. }

{Compared with oversampling techniques, KFRG has its advantages. One of the advantages is that KFRG is of good generalization ability and able to avoid overfitting problems.
In fact, randomly oversampling tends to suffer from overfitting problems as the minority samples in its training set were duplicated at random. As a result, the trained model becomes too specific to the training data and may not generalize well to new data. Other oversampling techniques are all based on SMOTE, which eliminate the harms of skewed distribution by creating new minority class samples. They generates a synthetic sample $x_{new}$ by using the linear interpolation of $x$ and $y$ with the expression of $x_{new} = x + (y-x)\times \alpha$,
where $\alpha$ is a random number in the range $[0,1]$ and $y$ is a $k$ nearest neighbour (kNN) of $x$ in the minority set. However, in many cases the kNN-based approach may generate wrong minority class samples as the above equation says that $x_{new}$  will lie in the line segment between $x$ and $y$. In addition, it would be difficult to find an appropriate
value of $k$ for kNN a priori. The parameter $k$ varies with the distribution of samples between minority and majority.
Different from data-level methods of oversampling, feature selection of KFGR neither duplicate nor generate any additional data. It keeps the distribution and class imbalance ratio of the data.}

{The other advantage is that the result of KFRG is interpretable. The selected features are the most distinguishing ones between pulsars and non-pulsars. In addition, KFRG perform well based on the data characteristics that regardless of the classifier used. Although the sets of selected features by KFRG may be changed with different classifiers, most of the selected features are the same, as it will be explained next section.}

\section{Discussion}
\label{Sec:discussion}
KFRG has been evaluated on HTRU. It proves that models based on KFRG achieve larger recall, precision $F_1$ score and less FPR than those without any feature selection. In other words, these selected features are distinguishable enough to pick out pulsar signals from the candidates.
The improvement of performance metrics comes for two reasons. For one thing, the Relief algorithm can filter out most of the irrelevant features from the feature pool. As we have explained, Relief score is based on the identification of feature value differences between nearest neighbor instance pairs with both the same class and different class. Thus, a feature with lower Relief score implies that the overlapping part of the two categories is large on the feature, which is considered as an irrelevant feature. For another, KFRG enables a classifier to select its most relevant features in a greedy way, as its objective function is to maximize the $F_1$ score of a given classifier.

Although the selected features by KFRG may be various for different classifiers, the most relevant features are almost the same.  The importance of features was evaluated by their frequency of being selected and ranked with stars according to the KFRG results (Table \ref{Tab:frequency}). Features M5 and L4 are three-star, as they are definitely selected by all the classifiers. Feature M3, M2, T1 and T5 are ranked as two-star, as they are chosen by about half of these classifiers. Features selected by only one or two classifiers are marked with one star. Those features left were never chosen by KFRG, implying they are redundant or irrelevant according to our experiments.

\begin{table*}[htbp]
\begin{center}
\caption{Importance of the selected features. The frequencies of the selected features of all the classifiers were counted according to Table \ref{Tab:greedy}. We use stars to rank the importance of features. }\label{Tab:frequency}
\setlength{\tabcolsep}{1.5mm}{
\begin{tabular}{lccccccc}
\noalign{\smallskip}
ID&Feature&Frequency& Importance \\
  \hline
L4&$Pf_{s}$    &	8/8&	 $\star \star \star$\\
M5&$\chi_{SNR}$&    7/8&	 $\star \star \star$\\
M3&$Log(P/DM)$ &	5/8&	 $\star \star $\\
M2&$D_{eq}$    &	4/8&	 $\star \star $\\
T1&SubbandCorr$_{\mu}$ &	4/8	&$\star\star$ \\
T5&SubintCorr$_{\mu}$ &	4/8	&$\star\star$ \\
M6&$D_{RMS}$   & 	2/8&	$\star$ \\
L3&$Pf_{k}$    &	1/8&	$\star  $\\
T6&SubintCorr$_{\sigma}$ &	1/8	&$\star$ \\
others&	0/8&	 &  \\
\hline
\end{tabular}
}
\end{center}
\end{table*}

{Notice that L4 and M5 are the most important features for most of the classifiers. They will be discussed in detail.
L4 is the other feature of the most relevant features for all the classifiers. It notes the skewness on the folded profile ($Pf_{s}$), which is a statistic value for the distribution of the pulse folded profile $P=\{p_i\}_{i=1}^{n}$, i.e.,
$$
{Pf_{s}=\frac{1}{n}\sum_{i=1}^{n}\left(\frac{p_i-\mu}{\sigma}\right)^{3}
=\frac{\frac{1}{n}\sum_{i=1}^{n}(p_i-\mu)^{3}}{\left(\frac{1}{n}\sum_{i=1}^{n}(p_i-\mu)^{2}\right)^{3 / 2}}}{,}
$$
{where, $\mu$, $\sigma$ are the mean and the standard deviation of $p_i$, respectively.}
{A candidate with large L4 implies that there is a great skewness of the folded profile. Skewness describes the symmetry of the distribution of a signal. A signal with a great skewness is likely a signal with a distinctly detectable pulse.

M5, standing for $\chi_{(SNR)}$, represents the persistence of the signal in the time domain, which is defined as the average score of $\chi_{(s)}$ \citep{morello2014spinn}, i.e., $\chi_{(SNR)} = \frac{1}{N}\sum_{i=1}^{N}\chi_{(s)}$, and}
$$
                                {\chi_{(s)}=}
                                \begin{cases}
                                1-exp(-\frac{s}{b}),  &s\geq 0,\\
                                \frac{s}{b},  &s<0,
                                \end{cases}$$
{where $s$ is the $SNR$ of the candidate in a sub-integration, and $b=\frac{16}{\sqrt{n_{\mathrm{sub}}}}$ presents the benchmark of the SNR, where $n_{\mathrm{sub}}$ is the total number of sub-integrations}.
The design basis of M5 is the fact that a genuine pulsar is expected to be consistently visible during most of an observation. As for man-made signals like RFIs, most of them last for a very short time, and then become invisible in a part of the observation. Therefore, M5 provides an effective selection criterion against these impulsive artificial signals.

  A scatter plot with coordinates of Feature M5 and L4 in Fig. \ref{5_vs_10} shows that: most of the non-pulsars can be easily separated from pulsars with these two features, as candidates with large L4 and M5 tend to be judged as pulsars. That could explain why they are frequently selected by most of the classifiers and very significant for the pulsar candidates sifting.
\begin{figure}
  \centering
  \includegraphics[width=14 cm]{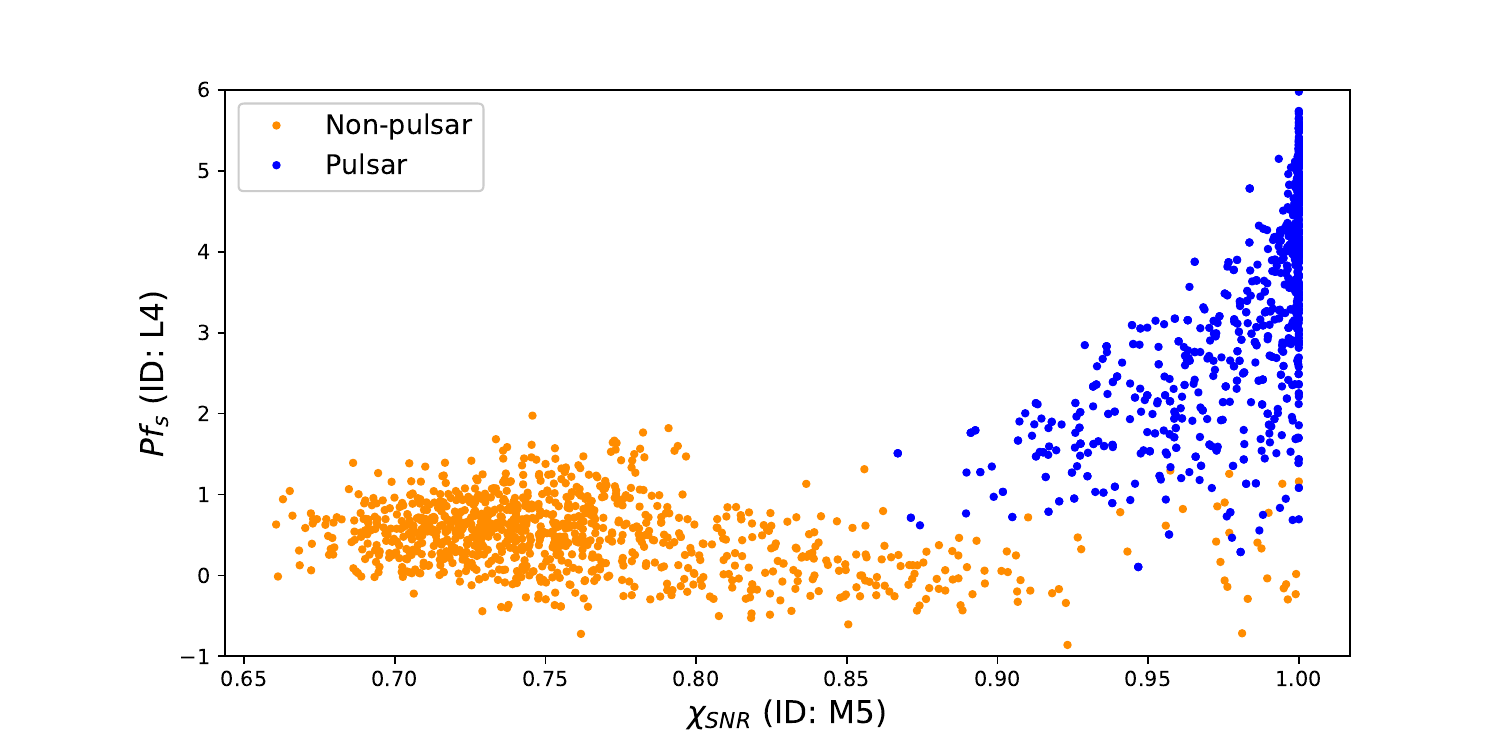}\\
  \caption{A scatter plot with coordinates of Feature M5 and L4.}\label{5_vs_10}
\end{figure}

In this work, a novel feature selection algorithm KFRG is proposed to improve the performance of PCS models in the class-imbalanced case.
KFRG combines Relief scores with Greedy algorithm to removes most of the redundant and irrelevant features.
Experiments based on HTRU show that KFRG is effective. Compared with models without any feature selection, the recall rate of the models based on KFRG features is higher and the $FPR$ is lower. Compared with some typical oversampling techniques, KFRG is more robust and interpretable besides better performance metrics.
Also, the importance of selected features by KFRG are described and explained in our work.

These experimental conclusions based on KFRG are efficient and practical, providing the potential guide to study machine learning methods for the candidate sifting, and serve other surveys of the next-generation radio telescopes.

\section*{Acknowledgements}
      Authors are grateful for support from the National Natural Science Foundation of China (grant No: 11973022,12373108), the Natural Science Foundation of Guangdong Province (No. 2020A1515010710), and Hanshan Normal University Startup Foundation for Doctor Scientific Research (No. QD202129).

\section*{Data availability}
The data underlying this article are publicly available in Centre for Astrophysics and Supercomputing, at {\url{h ttp://astronomy.swin.edu.au/~vmorello/}}, which was released by \cite{morello2014spinn}. The detailed description of the data is in Section \ref{sec:dataset}.



\bibliographystyle{raa}
\bibliography{references}

\begin{thebibliography}{57}
\providecommand\natexlab[1]{#1}
\providecommand\JournalTitle[1]{#1}

\bibitem[Baghram {et~al.}(2011)]{baghram2011prospects}
Baghram, S., Afshordi, N., \& Zurek, K.~M. 2011, Physical Review D, 84, 043511

\bibitem[Bates {et~al.}(2012)]{bates2012high}
Bates, S., Bailes, M., Barsdell, B., {et~al.} 2012, Monthly Notices of the
  Royal Astronomical Society, 427, 1052

\bibitem[Bethapudi \& Desai(2018)]{bethapudi2018separation}
Bethapudi, S., \& Desai, S. 2018, Astronomy and computing, 23, 15

\bibitem[Black(2012)]{black2012greedy}
Black, P.~E. 2012, US Nat. Inst. Std. \& Tech Report, 88, 95

\bibitem[Chawla {et~al.}(2004)]{2004SPECIAL}
Chawla, N., Japkowicz, N., \& Kolcz, A. 2004

\bibitem[Chawla {et~al.}(2002)]{chawla2002smote}
Chawla, N.~V., Bowyer, K.~W., Hall, L.~O., \& Kegelmeyer, W.~P. 2002, Journal
  of artificial intelligence research, 16, 321

\bibitem[Chen \& Guestrin(2016)]{chen2016xgboost}
Chen, T., \& Guestrin, C. 2016, in Proceedings of the 22nd acm sigkdd
  international conference on knowledge discovery and data mining, ACM, 785

\bibitem[Devine {et~al.}(2016)]{devine2016detection}
Devine, T.~R., Goseva-Popstojanova, K., \& McLaughlin, M. 2016, Monthly Notices
  of the Royal Astronomical Society, 459, 1519

\bibitem[Dietterich {et~al.}(2002)]{dietterich2002ensemble}
Dietterich, T.~G., {et~al.} 2002, The handbook of brain theory and neural
  networks, 2, 110

\bibitem[Eatough {et~al.}(2010)]{eatough2010selection}
Eatough, R.~P., Molkenthin, N., Kramer, M., {et~al.} 2010, Monthly Notices of
  the Royal Astronomical Society, 407, 2443

\bibitem[Freund \& Schapire(1997)]{freund1997decision}
Freund, Y., \& Schapire, R.~E. 1997, Journal of computer and system sciences,
  55, 119

\bibitem[Goldreich(2008)]{goldreich2008computational}
Goldreich, O. 2008, ACM Sigact News, 39, 35

\bibitem[Guo {et~al.}(2019)]{guo2019pulsar}
Guo, P., Duan, F., Wang, P., {et~al.} 2019, Monthly Notices of the Royal
  Astronomical Society, 490, 5424

\bibitem[Gupta(1960)]{gupta1960point}
Gupta, S.~D. 1960, Psychometrika, 25, 393

\bibitem[Guyon \& Elisseeff(2003)]{guyon2003introduction}
Guyon, I., \& Elisseeff, A. 2003, Journal of machine learning research, 3, 1157

\bibitem[Han {et~al.}(2005)]{2005Borderline}
Han, H., Wang, W.~Y., \& Mao, B.~H. 2005, Lecture Notes in Computer Science

\bibitem[Han {et~al.}(2004)]{han2004spatial}
Han, J.-L., Ferriere, K., \& Manchester, R. 2004, The Astrophysical Journal,
  610, 820

\bibitem[Hastie {et~al.}(2005)]{hastie2005elements}
Hastie, T., Tibshirani, R., Friedman, J., \& Franklin, J. 2005, The
  Mathematical Intelligencer, 27, 83

\bibitem[Haykin(1994)]{haykin1994neural}
Haykin, S. 1994, Neural networks: a comprehensive foundation (Prentice Hall
  PTR)

\bibitem[He {et~al.}(2008)]{he2008adasyn}
He, H., Bai, Y., Garcia, E.~A., \& Li, S. 2008, in 2008 IEEE international
  joint conference on neural networks (IEEE world congress on computational
  intelligence), Ieee, 1322

\bibitem[Hosmer~Jr {et~al.}(2013)]{hosmer2013applied}
Hosmer~Jr, D.~W., Lemeshow, S., \& Sturdivant, R.~X. 2013, Applied logistic
  regression, Vol. 398 (John Wiley \& Sons)

\bibitem[Japkowicz {et~al.}(2000)]{japkowicz2000learning}
Japkowicz, N., {et~al.} 2000, in AAAI workshop on learning from imbalanced data
  sets, Vol.~68, AAAI Press Menlo Park, CA, 10

\bibitem[Keith {et~al.}(2010)]{keith2010high}
Keith, M., Jameson, A., Van~Straten, W., {et~al.} 2010, Monthly Notices of the
  Royal Astronomical Society, 409, 619

\bibitem[Kira \& Rendell(1992)]{kira1992practical}
Kira, K., \& Rendell, L.~A. 1992, in Machine Learning Proceedings 1992
  (Elsevier), 249

\bibitem[Levin {et~al.}(2013)]{levin2013high}
Levin, L., Bailes, M., Barsdell, B., {et~al.} 2013, Monthly Notices of the
  Royal Astronomical Society, 434, 1387

\bibitem[Liaw {et~al.}(2002)]{liaw2002classification}
Liaw, A., Wiener, M., {et~al.} 2002, R news, 2, 18

\bibitem[Lin {et~al.}(2020)]{lin2020pulsar}
Lin, H., Li, X., \& Zeng, Q. 2020, The Astrophysical Journal, 899, 104

\bibitem[Lyon {et~al.}(2014)]{lyon2014hellinger}
Lyon, R., Brooke, J., Knowles, J., \& Stappers, B. 2014, in 2014 22nd
  International Conference on Pattern Recognition, IEEE, 1969

\bibitem[Lyon {et~al.}(2016)]{lyon2016fifty}
Lyon, R.~J., Stappers, B., Cooper, S., Brooke, J., \& Knowles, J. 2016, Monthly
  Notices of the Royal Astronomical Society, 459, 1104

\bibitem[Mahmoud \& Guo(2021)]{mahmoud2021learning}
Mahmoud, M.~A., \& Guo, P. 2021, New Astronomy, 85, 101561

\bibitem[Maldonado {et~al.}(2014)]{2014Feature}
Maldonado, S., Weber, R., \& Famili, F. 2014, Information Sciences, 286, 228

\bibitem[Manchester {et~al.}(2001)]{manchester2001parkes}
Manchester, R.~N., Lyne, A.~G., Camilo, F., {et~al.} 2001, Monthly Notices of
  the Royal Astronomical Society, 328, 17

\bibitem[Mitchell {et~al.}(1997)]{mitchell1997machine}
Mitchell, T.~M., {et~al.} 1997, Burr Ridge, IL: McGraw Hill, 45, 870

\bibitem[Mohri {et~al.}(2018)]{mohri2018foundations}
Mohri, M., Rostamizadeh, A., \& Talwalkar, A. 2018, Foundations of machine
  learning (MIT press)

\bibitem[M{\"o}ller {et~al.}(2016)]{moller2016photometric}
M{\"o}ller, A., Ruhlmann-Kleider, V., Leloup, C., {et~al.} 2016, Journal of
  Cosmology and Astroparticle Physics, 2016, 008

\bibitem[Morello {et~al.}(2014)]{morello2014spinn}
Morello, V., Barr, E., Bailes, M., {et~al.} 2014, Monthly Notices of the Royal
  Astronomical Society, 443, 1651

\bibitem[Nan(2006)]{nan2006five}
Nan, R. 2006, Science in China series G, 49, 129

\bibitem[Nan {et~al.}(2016)]{nan2016fast}
Nan, R., Zhang, H., Zhang, Y., {et~al.} 2016, Acta Astronomica Sinica, 57, 623

\bibitem[Nan {et~al.}(2011)]{nan2011five}
Nan, R., Li, D., Jin, C., {et~al.} 2011, International Journal of Modern
  Physics D, 20, 989

\bibitem[Quinlan(2014)]{quinlan2014c4}
Quinlan, J.~R. 2014, C4. 5: programs for machine learning (Elsevier)

\bibitem[Ransom(2001)]{ransom2001new}
Ransom, S.~M. 2001, New search techniques for binary pulsars, Vol. 119 (Harvard
  University Harvard, USA)

\bibitem[Shannon(1948)]{shannon1948mathematical}
Shannon, C.~E. 1948, Bell system technical journal, 27, 379

\bibitem[Smits {et~al.}(2009)]{smits2009pulsar}
Smits, R., Kramer, M., Stappers, B., {et~al.} 2009, Astronomy \& Astrophysics,
  493, 1161

\bibitem[Suykens \& Vandewalle(1999)]{suykens1999least}
Suykens, J.~A., \& Vandewalle, J. 1999, Neural processing letters, 9, 293

\bibitem[Tan {et~al.}(2017)]{tan2017ensemble}
Tan, C., Lyon, R., Stappers, B., {et~al.} 2017, Monthly Notices of the Royal
  Astronomical Society, 474, 4571

\bibitem[Tang {et~al.}(2014)]{tang2014feature}
Tang, J., Alelyani, S., \& Liu, H. 2014, Data classification: Algorithms and
  applications, 37

\bibitem[Taylor~Jr(1994)]{taylor1994binary}
Taylor~Jr, J.~H. 1994, Reviews of Modern Physics, 66, 711

\bibitem[Urbanowicz {et~al.}(2018)]{urbanowicz2018relief}
Urbanowicz, R.~J., Meeker, M., La~Cava, W., Olson, R.~S., \& Moore, J.~H. 2018,
  Journal of biomedical informatics, 85, 189

\bibitem[van Haarlem {et~al.}(2013)]{van2013lofar}
van Haarlem, M.~{\'a}., Wise, M., Gunst, A., {et~al.} 2013, Astronomy \&
  astrophysics, 556, A2

\bibitem[Wang {et~al.}(2019)]{wang2019pulsar}
Wang, H., Zhu, W., Guo, P., {et~al.} 2019, Science China Physics, Mechanics \&
  Astronomy, 62, 1

\bibitem[Xiao-fei {et~al.}(2021)]{XIAOFEI2021364}
Xiao-fei, L., Bao-qiang, L., Tao, A., Zhi-jun, X., \& Zhong-li, Z. 2021,
  Chinese Astronomy and Astrophysics, 45, 364

\bibitem[Xiao {et~al.}(2020)]{xiao2020pulsar}
Xiao, J., Li, X., Lin, H., \& Qiu, K. 2020, Monthly Notices of the Royal
  Astronomical Society, 492, 2119

\bibitem[Yin {et~al.}(2013)]{2013Feature}
Yin, L., Yong, G., Xiao, K., Wang, X., \& Quan, X. 2013, Neurocomputing, 105, 3

\bibitem[Yuanyu {et~al.}(2019)]{Yuanyu2019A}
Yuanyu, He, Junhai, {et~al.} 2019, Computational Biology and Chemistry

\bibitem[Zeng {et~al.}(2020)]{zeng2020concat}
Zeng, Q., Li, X., \& Lin, H. 2020, Monthly Notices of the Royal Astronomical
  Society, 494, 3110

\bibitem[Zhang {et~al.}(2019)]{zhang2019pulsar}
Zhang, H., Zhao, Z., An, T., Lao, B., \& Chen, X. 2019, Computers \& Electrical
  Engineering, 73, 1

\bibitem[Zhu {et~al.}(2014)]{zhu2014searching}
Zhu, W., Berndsen, A., Madsen, E., {et~al.} 2014, The Astrophysical Journal,
  781, 117

\end{thebibliography}

\label{lastpage}

\end{document}